\documentclass{osa-article}
\usepackage{threeparttable}
\usepackage{siunitx}
\usepackage{graphicx}
\usepackage{multirow}
\journal{oe}

\articletype{Research Article}

\begin{document}

\title{Planar multi-aperture fish-eye lens using metagrating}

\author{Zihan Zang,\authormark{1,5} Haoqiang Wang,\authormark{2,5} Yanjun Han,\authormark{2,3,4}\\ Hongtao Li,\authormark{2} {H.Y. Fu},\authormark{1} and Yi Luo\authormark{2,3,4,6}}

\address{\authormark{1}Tsinghua-Berkeley Shenzhen Institute (TBSI), Tsinghua University, Shenzhen 518055, China\\
	\authormark{2}Beijing National Research Centre for Information Science and Technology, Department of Electronic Engineering, Tsinghua University, Beijing 10084, China\\
	\authormark{3}Center for Flexible Electronics Technology, Tsinghua University, Beijing 100084, China\\
	\authormark{4}Flexible Intelligent Optoelectronic Device and Technology Center, Institute of Flexible Electronics Technology of THU, Zhejiang, Jiaxing 304006, China\\
	\authormark{5}These authors contributed equally to this work.}

\email{\authormark{6}luoy@tsinghua.edu.cn}



\begin{abstract}
	The design of compact optical systems with large field of view has been difficult due to the requirement of many elements or a curved focal plane to reduce off-axis aberration. We propose a multi-aperture lens design to effectively resolve these issues. Metagrating-based deflectors are placed near entrance pupils of multi-aperture lens array to enhance field of view. A systematic design method is given in details. In design examples, a \ang{+-80} field of view using only two planar optical elements is achieved. Also, the system is extremely compact with total track lengths an order of magnitude smaller than conventional fish-eye lenses, while the imaging performance is comparable with conventional designs.
\end{abstract}

\section{Introduction}

\begin{figure}[ht!]
	\centering\includegraphics[width=13cm]{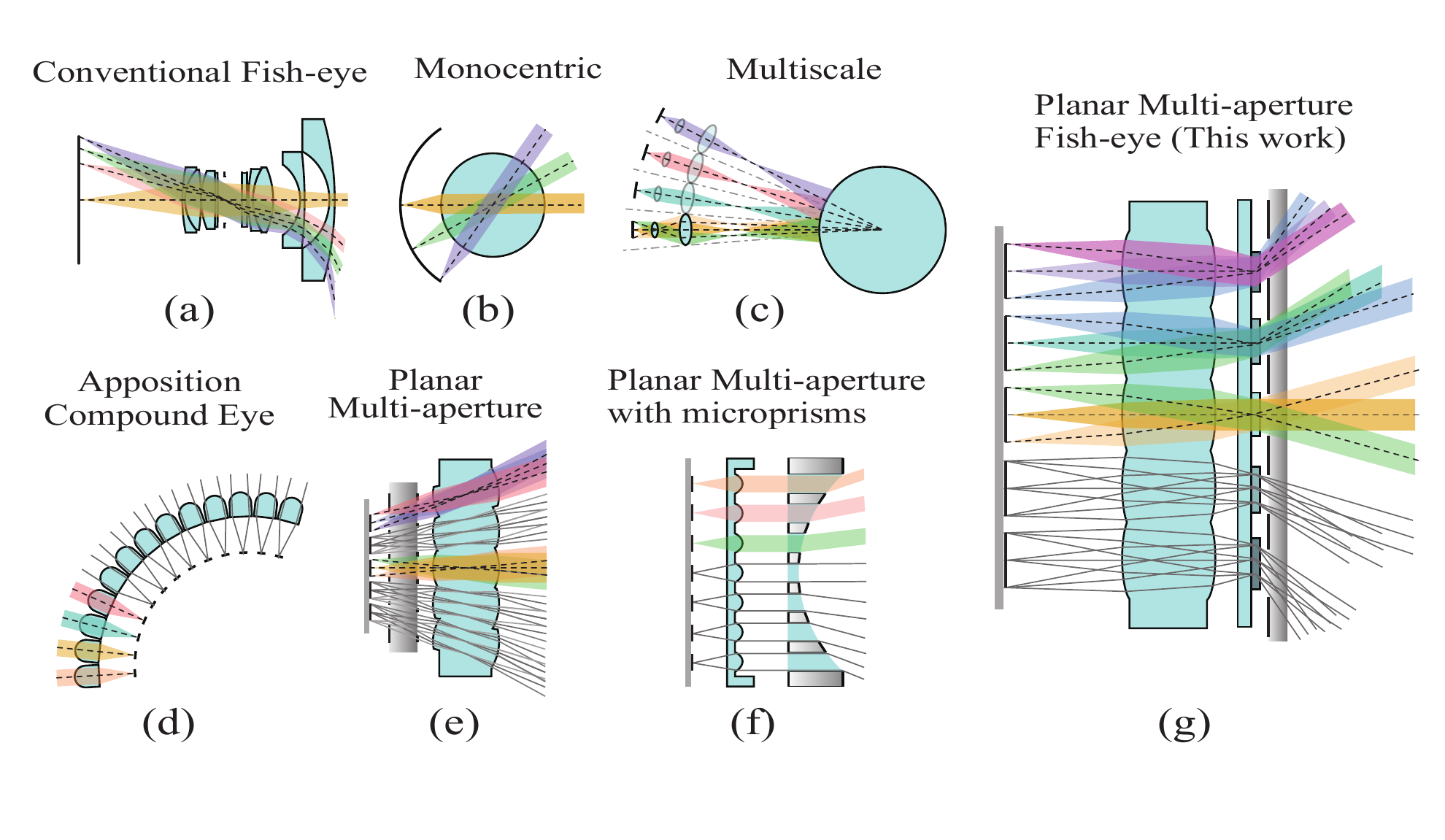}
	\caption{Lenses with large FoV (not to scale). (a) conventional fish-eye lens with many elements (adopted from Zebase library (F016)); (b) monocentric lens with a spherical focal plane; (c) multiscale lens with a series of discrete focal planes aligned with a sphere; (d) bio-inspired elastomeric lens based on apposition compound eye with an curved structure; (e) planar multi-aperture lens with a limited FoV; (f) planar multi-aperture lens with microprisms, also with a limited FoV; (g) the proposed planar multi-aperture fish-eye lens using metagrating array, with a large FoV and a planar focal plane.}
	\label{gallery}
\end{figure}

Imaging system often prefers a large field of view (FoV) to acquire more information. In the context of surveillance camera, light detection and ranging (LiDAR) and tools for endoscopy, field performance is especially important and a FoV of $\pm 90^{\circ}$ is often required \cite{yogamani2019woodscape,kim2015surveillance,qureshi2006surveillance,etsuko2000endoscope}. Such lenses are often referred as fish-eye lenses. Optical systems with such a large FoV suffer from large off-axis aberration including astigmatism, field curvature and coma. In conventional optical design, multiple lenses are combined to perform correction and a bulky, very strongly curved negative front element is usually needed (see Fig. \ref{gallery}(a)). In most cases, 6 to 10 or even more elements are needed \cite{Kumler2000}, which makes it difficult to achieve a compact and low-cost design. Moreover, off-axis aberrations usually cannot be fully removed, especially at large field angles. Therefore, imaging quality decreases with field angle and total achievable resolution points or space-bandwidth product is limited.

Many methods have been proposed to reduce off-axis aberration and expand the FoV of optical systems \cite{Kingslake1989,Rim2008,Cossairt2011,Stamenov2014,Brady2009,Tremblay2012,Pang2018,Brady2018,song2013}. Monocentric lens is the earliest but powerful one among them \cite{Kingslake1989}. As shown in Fig. \ref{gallery}(b), monocentric lenses have a highly symmetric structure with front and back surfaces sharing a common center, which offers on-axis performance for all field angles. For example, a simple ball lens with a curved image surface can produce a superior image within \ang{+-90} FoV \cite{Cossairt2011}. However, monocentric design highly relies on a curved focal plane \cite{Rim2008}, and thus requires a curved imager, which is extremely challenging to fabricate. Therefore, monocentric design has rarely been adopted, except for some large-scale camera arrays \cite{Cossairt2011}. Recently, a fiber bundle with curved end surfaces is used to couple the curved image to a common planar imager so that a planar imager can be implemented in a monocentric design \cite{Stamenov2014}. Although this solution is practical, the use of fiber causes coupling problems and makes the system complex and bulky.

In 2009, Brady \textit{et al.} proposed a heterogeneous method or multiscale lens design \cite{Brady2009}. As shown in Fig. \ref{gallery}(c), it can be treated as a combination of multi-aperture strategy and monocentric design, where a multi-aperture lens is placed behind a single-aperture objective lens, such as a ball lens. The introduce of objective lens enhances the capability of off-axis-aberration reduction at large field angles and a large FoV and diffraction limited imaging quality can be achieved \cite{Tremblay2012,Pang2018}. However, the requirement of an overall curved focal plane and the complex structure hinder miniaturization. Hence, multiscale lens design is mainly built for large-scale camera arrays\cite{Brady2018}. To enable a miniaturized design, elastomeric cameras inspired by apposition compound eye of arthropod (as shown in Fig. \ref{gallery}(d)) is demonstrated with an FoV of \ang{+-80} \cite{song2013}. However, the drawback of the compound-eye structure is the severely limited spatial resolution. This demonstration only possesses a total pixel count of 166. Besides, the fabrication process is also challenging because of the curved alignment structure.

To alleviate the difficulty of fabrication and enable miniaturization or even wafer-level camera, a design with planar focal plane is highly preferred. However, a planar focal plane contradicts the expansion of FoV. A compromise solution is multi-aperture design, which is exhibited in Fig. \ref{gallery}(e). The basic idea behind multi-aperture design is similar to the aforementioned methods: with FoV segmentation, a large FoV is split into multiple smaller sub-FoVs, which can be processed separately. For this type of design, off-axis aberration is reduced and optical design can be optimized locally rather than globally, which simplified the overall structure. Nevertheless, as off-axis aberration cannot be effectively reduced in a planar multi-aperture lens, only relatively small FoV (such as \ang{+-35}) can be achieved \cite{Bruckner2011,kim2020}. To improve the ability of off-axis aberration reduction of a multi-aperture lens and further simplify the structure, microprism array has been introduced to bend the sub-FoVs\cite{keum2018} (see Fig. \ref{gallery}(f)). However, microprisms can only offer limited bending ability and the system only shows a FoV of \ang{+-34}, which is still far from fish-eye lenses.

Recently, flat optics technology based on ultra-thin optical scatter arrays to shape wavefronts has been extensively studied, such as metasurface and diffractive elements. Flat diffractive lens and metalens are emerging optical components to achieve miniaturization \cite{banerji2019}. Especially, ultrathin single-element fish-eye metalens has been reported recently\cite{shalaginov2020}. However, the global optimization of such metalens is challenging for the large design area\cite{Phan2019}, which decreases the overall efficiency. Moreover, the fish-eye metalens design suffers from large Fresnel loss at large field angles because of flat entrance window. On the other hand, metasurface-based deflector, i.e. metagrating, shows superior efficiency and relatively simple design process. Ultra-thin metagratings with large bending angle up to \ang{85}, high efficiency up to $90\%$ and polarization-independent characteristic have been reported \cite{Sell2017, Sell2018, Qiao2018, Neder2019, Dong2019, Jiang2020}. However, the potential application of such metagratings has not been well studied.

\begin{figure}[ht!]
	\centering\includegraphics[width=5cm]{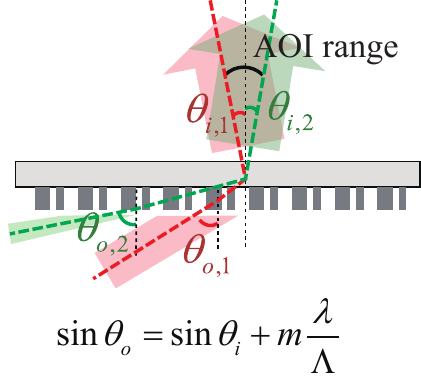}
	\caption{Metagratings and the acceptable AOI range. Note that AOI is defined from the side with smaller angle, even if the input beam is from another side.}
	\label{metagrating}
\end{figure}

In this paper, we propose to combine a metagrating array and a multi-aperture lens to fully eliminate the need of curved focal planes and facilitate a more compact, high-efficienty design for fish-eye optical system (see Fig. \ref{gallery}(g)). The metagrating array contains a series of metagrating units, which can efficiently bend light with large angles. Note that for each metagrating unit, the range of angle of incidence (AOI) is limited by FoV segmentation offered by the multi-aperture lens, which is usually in the range of 5 to 15 degrees \cite{Sell2018,zotero-280}. 
Here, AOI is defined from the side with smaller angle to be consistent with literature above, which is show in Fig. \ref{metagrating}. However, the actual input beam of the proposed system is from larger angle side. The relation bewtween AOI $\theta_i$ and output angle $\theta_o$ is given by grating equation (also shown in the figure), where $\Lambda$ is grating period; $\lambda$ is input wavelength and $m$ is diffraction order. Each lenslet can be separately optimized, which is another key merit to achieve compactness. First-order design of the system is given in Section 2, which reveals some basic limitation on system parameters. Based on the analysis on spatial resolution and scaling performance given in Section 3, the parameters can be determined. Our results show that the achievable FoV can be greatly improved while keeping a compact design and a high resolution at large field angle. Two typical design examples with \ang{+-80} FoV are given in Section 4, which are balanced between spatial resolution and total track length. These examples show a comparable performance with conventional fish-eye lenses with many elements. Besides, the systems are very compact with a total track length an order of magnitude smaller than that of conventional design.

\section{First-order design}

\begin{figure}[ht!]
	\centering\includegraphics[width=5.5cm]{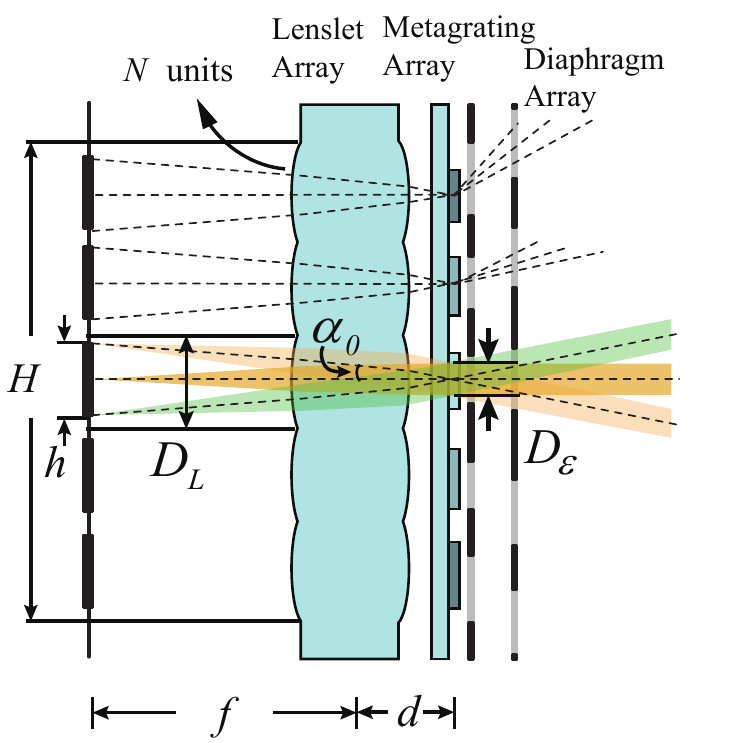}
	\caption{First-order model. $f$, focal length of each lenslet; $d$ separation distance between lenslet array and metagrating array; $N$, unit number; $H$, lateral imager height; $h$, the height of a sub-image formed by a single lenslet; $D_L$, unit diameter; $D_\varepsilon$, entrance pupil diameter; $\alpha_0$, half sub-FoV of each lenslet, which is also the range of angle of incidence (AOI) of metagrating units}
	\label{structure}
\end{figure}

\begin{figure}[ht!]
	\centering\includegraphics[width=0.9\textwidth]{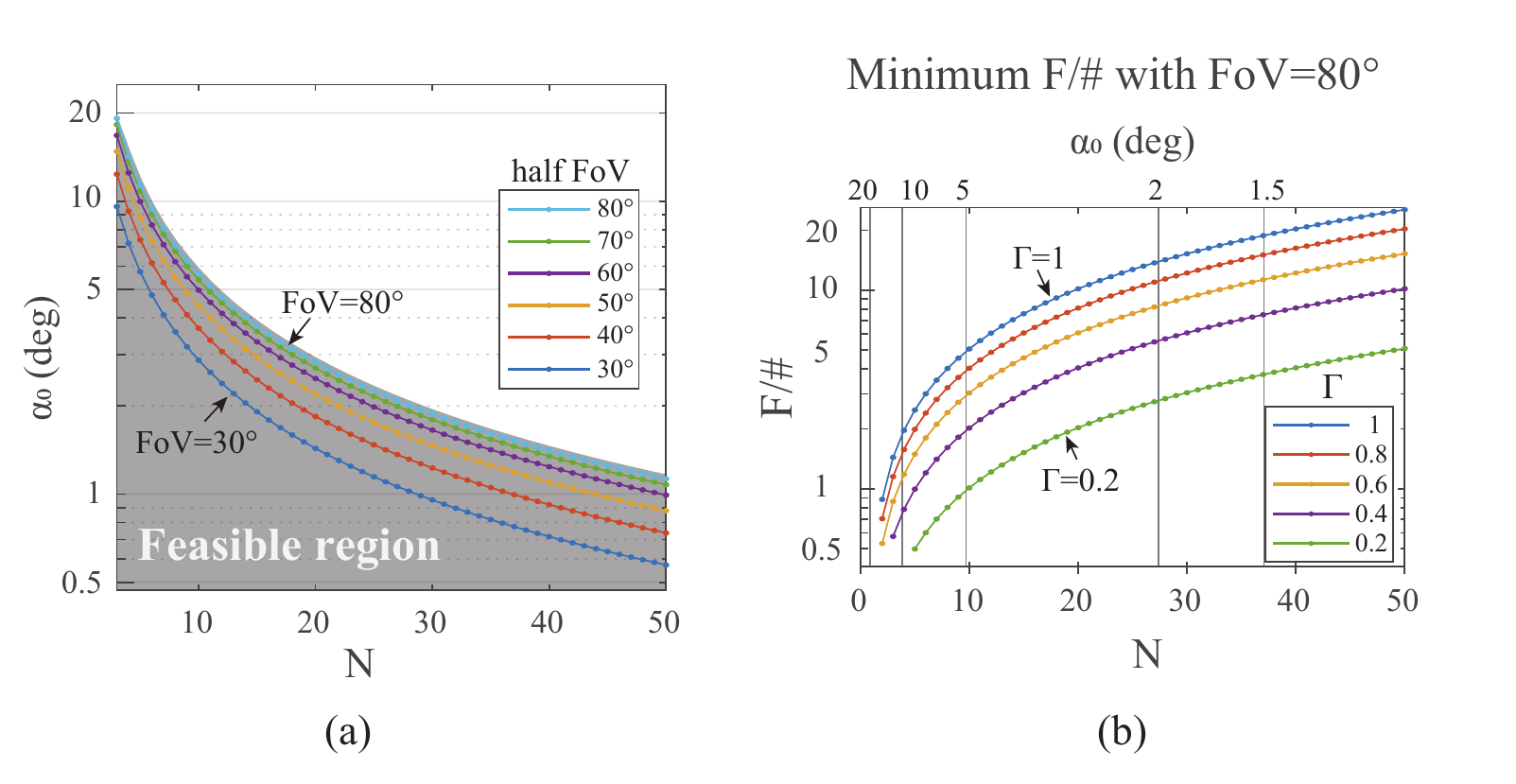}
	\caption{(a) Required metagratig AOI range $[-\alpha_0,\alpha_0]$ and number of units $N$ to achieve a system half FoV. (b) Minimum achievable F number when different $\alpha_0$ and $N$ are chosen to achieve a system half FoV of \ang{80}}
	\label{fov}
\end{figure}

The proposed design contains a lenslet array and a metagrating array. To surpress stray light, two diaphragm array are added. For simplicity, first-order model is created as shown in Fig. \ref{structure}, an one-dimensional (1D) array with $N$ units and total height $H$ is considered here. Each unit has a lateral length $D_L$ and is able to form a sub-image with height $h$. As the size of sub-images can be less than unit size, we define their ratio as fill factor $\Gamma$ ranging from 0 to 1, i.e.,
\begin{equation}
	\Gamma = \frac{h}{D_L}=\frac{N h}{H}, \;\;0<\Gamma \leq 1.
	\label{fill}
\end{equation}
Then we set the maximum field angle $\alpha_0$ of each lenslet unit. Note that $[-\alpha_0,\alpha_0]$ now becomes the operating AOI range of all metagrating units, which is limited by metagrating design feasibility. Once $\alpha_0$ is decided, the focal length $f$ of the lenslet array is given by
\begin{equation}
	f=\frac{h}{2\tan \alpha_0}=\frac{\Gamma H}{2N\tan \alpha_0},
	\label{focal_len}
\end{equation}
For a metagrating with grating period $\Lambda$, a incident beam with wavelength $\lambda$ and AOI $\theta_i$ will be deflected to an output angle $\theta_o$, which is given by the grating equation:
\begin{equation}
	\sin\theta_o = \sin\theta_i + m \frac{\lambda}{\Lambda},
\end{equation}
where $m$ is the operating order.
When AOI $\theta_i$ varies in the range of $[-\alpha_0,\ \alpha_0]$, the output angle $\theta_o$ will cover the range of $[\arcsin (g_m + \sin\alpha_0 ),\ \arcsin (g_m - \sin\alpha_0 )]$, where $g_m = m \lambda/\Lambda$ is the length of grating vector. The range of the output angle is thus the sub-FoV of a single unit. Then the fully stitched half FoV of all the N units is given by
\begin{equation}
	\text{FoV} =  \arcsin (N \sin \alpha_0).
	\label{fov_eq}
\end{equation}
The length of the grating vector of the $j$-th metagrating unit counted from the center position is given by
\begin{equation}
	g_{m,j} =\frac{m_j\lambda}{\Lambda_j}=
	\begin{cases}
		2j\sin \alpha_0,     & N \text{ is odd, } i =0,...,\frac{N-1}{2};\, j=0 \text{ for center unit}     \\
		(2j-1)\sin \alpha_0, & N \text{ is even, } i =1,...,\frac{N}{2};\, j=1\text{ for unit near center},
	\end{cases}
	\label{grating_lines}
\end{equation}
where $m_j$ and $\Lambda_j$ is the operating diffraction order and the grating period of $j$-th unit, respectively.
Using above relations, the system half FoV is decided by number of units $N$ and metagrating acceptable AOI range $[-\alpha_0,\alpha_0]$ as is shown in Fig. \ref{fov}(a).
Considering the fabrication difficulties and optical efficiency of the metagrating array, AOI range cannot be chosen too large. However, in most cases the required smaller AOI range can be easily fulfilled by using more units. From Fig. 4(a), we can find that if the AOI range is larger than \ang{2}, less than 30 units can cover up to \ang{+-80} FoV.

Next, we need to decide the required F number (which is defined by the ratio between focal length and entrance pupil diameter, i.e. $ \text{F/\#} = f/D_\varepsilon$) of each lenslet. To improve light collection efficiency, the diameter of entrance pupil $D_\varepsilon$ should be maximized and F/\# is thus minimized. However, non-overlap condition for metagrating units requires $D_\varepsilon<H/N$. Combining the realation with Eq.\ref{focal_len}, we see minimum F/\# of each lenslet is limited by both the AOI range of metagratings and fill factor of sub-images:
\begin{equation}
	\text{F/\#} \ge \frac{\Gamma}{2\tan \alpha_0}.
	\label{F_num}
\end{equation}
Using the relation in Eq. \ref{fov_eq}, we can express $2\tan\alpha_0$ as a function of $N$ at a certain target system half FoV:
\begin{equation}
\begin{aligned}
    2\tan\alpha_0 = f_{\text{FoV}}(N) &= 2\tan{(\arcsin{(\sin{(\text{FoV})}/N)})}\\ &\approx 2\sin{(\text{FoV})}/N,
\end{aligned}
\end{equation}
where the approximation holds when $\alpha_0$ is not large. 
Then Eq. \ref{F_num} can be written as
\begin{equation}
	\text{F/\#} \ge \frac{\Gamma}{f_{\text{FoV}}(N)} \approx \frac{\Gamma N}{2\sin{(\text{FoV})}}.
	\label{F_num2}
\end{equation}
With a fixed design FoV, smaller fill factor $\Gamma$ or unit number $N$ must be used to obtain smaller F/\# for maximum optical collection efficiency. For a target system FoV of \ang{+-80}, the relation between minimum achievable F/\#, $\Gamma$ and number of units $N$ is shown in Fig. \ref{fov}(b) by using Eq. \ref{F_num2}. To further decide F/\#, $\Gamma$ and $N$, analysis of spatial resolution is needed, which will be discussed in Section 3.    

Finally we consider the requirement of zero vignetting, which further limits the entrance aperture by
\begin{equation}
	\begin{split}
		D_\varepsilon &<D_L -2d\cdot\tan\alpha_0 \\
		&=H/N -2d\cdot\tan\alpha_0 ,
	\end{split}
	\label{vig}
\end{equation}
in which $d$ is the separation distance between the lenslet array and the metagrating array. This condition can be easily imposed during optic design process.

\section{Spatial resolution and scaling}
\begin{figure}[ht!]
	\centering\includegraphics[width=0.9\textwidth]{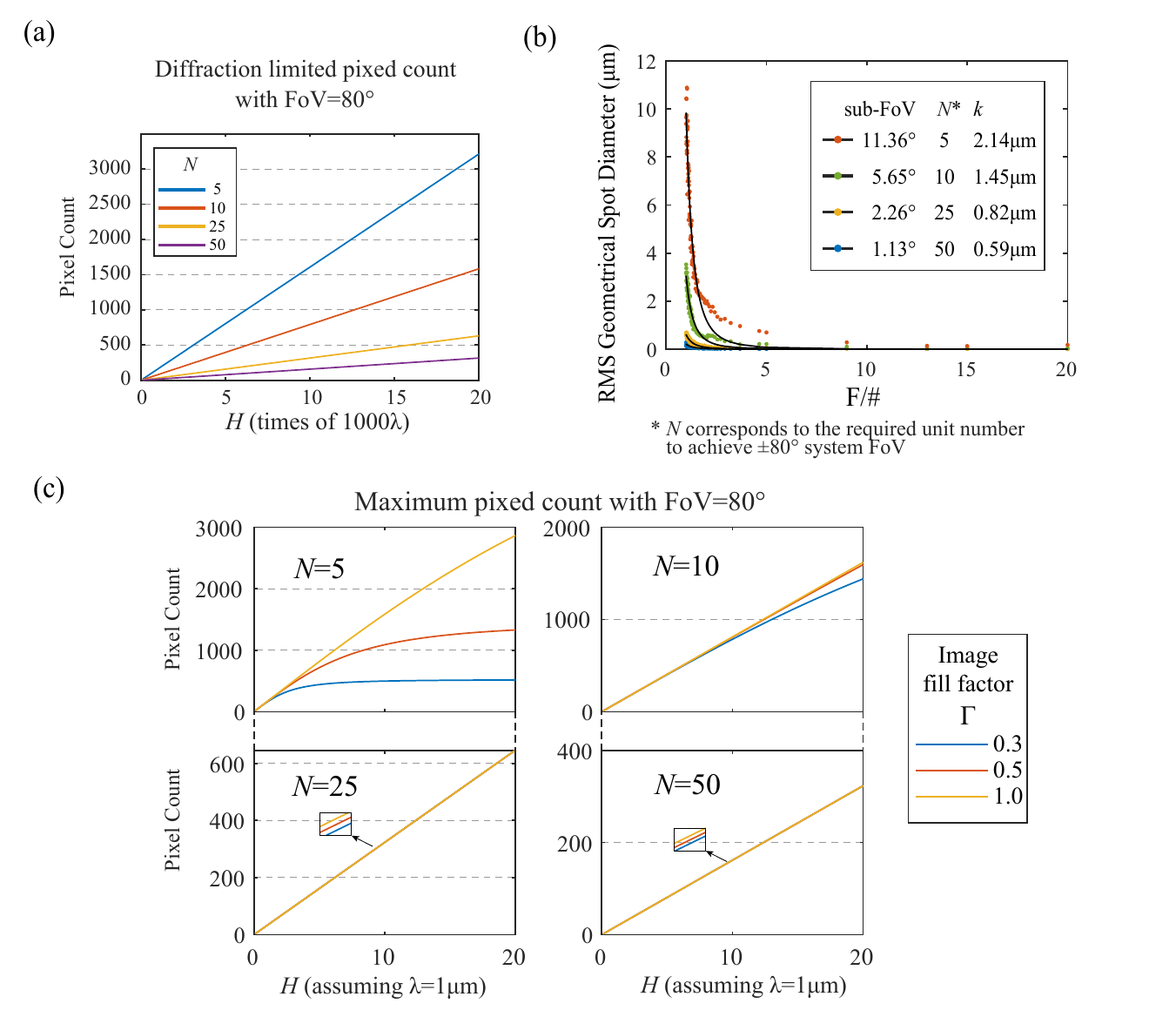}
	\caption{(a) geometric spot diameter and cubic fitting curve of a single lenslet with different sub-FoV and F/\#. E48R material and 4-th order aspherical surface are used. (b) Scaling performance of the total pixel count when using different unit number $N$. The target system FoV is \ang{+-80}. Only diffraction limit is considered. $H$, lateral imager height (see. Fig. \ref{structure}). (c)  Scaling performance of total pixel count when using different unit number $N$ and image fill factor $\Gamma$  (see Fig. \ref{structure}). Both geometric aberration and diffraction limit are included. $H$, lateral imager height.}
	\label{scaling}
\end{figure}
Using Rayleigh criterion, minimum achievable F/\# gives the diffraction limited spot size of the proposed system as $\delta_\text{d} = 2.44\lambda\cdot \text{F/\#}$. Therefore, the diffraction limited lateral pixel counts of an optical system is given by
\begin{equation}
	N_{\text{d}} = \frac{\Gamma H}{\delta_\text{d}} = 0.8\frac{H}{\lambda}\tan \alpha_0=0.4\frac{H}{\lambda}f_{\text{FoV}}(N),
\end{equation}
which provides the scaling performance of lateral pixel counts. The relation between the lateral imager height $H$ with the diffraction limited pixel count $N_d$ is illustrated in Fig. \ref{scaling}(a), which shows a linear relationship and can be seen as the upper limit of pixel count of the system.

To further consider geometric aberration and obtain a more accurate estimation of pixel count performance, estimation of geometric spot size $\delta_\text{g}$ is needed. Then the maximum lateral pixel counts becomes \cite{Lohmann1989}
\begin{equation}
	N_{\text{d,g}} = \frac{\Gamma H}{\sqrt{\delta_\text{d}^2+\delta_\text{g}^2}}.
	\label{cnt}
\end{equation}
As each lenslet unit is operated at relatively small sub-FoV, spherical aberrations will remarkable, leading to a empirical formula of geometric spot size \cite{Cossairt2011}:
\begin{equation}
	\delta_g = kM\frac{1}{\text{F/\#}^3}= k\frac{f}{f_0}\frac{1}{\text{F/\#}^3}= k\frac{H/N}{f_0}\frac{1}{\text{F/\#}^2},
	\label{delta_g}
\end{equation}
where $k$ is a constant decided by the shape and refraction index of a lenslet, and $f_0$ is a reference focal length which equals to 1 mm. If the system is scaled up by a factor of $M$, then the focal length of lenslets $f$ will be $M$ times larger, leading to a $M$-times larger geometric spot. Note that F/\# and FoV keep unchanged during lens scaling. To check the effectiveness of the formula, a series of lenslets with a range of F/\# and sub-FoV are optimized using an optical design software. The material is chosen as E48R (optical plastic) and up to 4-th order aspherical surface is used. The average geometric spot diameter of these optimized lenslets are shown in Fig. \ref{scaling}(b). Here, we select 4 kinds of lenslets which are suitable to apply in \ang{80}-FoV systems with 5, 10, 25 and 50 units. Least-square fitting is applied to evaluate the constant $k$ of each lenslet, and estimated pixel count can be calculated by Eq. \ref{cnt}. The scaling performance of estimated pixel count is illustrated in Fig. \ref{scaling}(c), which shows that diffraction limit dominates at small imager height, leading to a near-linear relation. However, geometric aberration will gradually dominate as imager height increases, leading to a deviation from linear relation. This effect is more remarkable for small unit number $N$ and small image fill factor $\Gamma$ conditions, which lead to a small F/\# and large geometric aberration. To select suitable scaling parameters, we need to maximize pixel count while keep the lateral imager height reasonable. As a result, unit number $N$ should not be too large. From Fig. \ref{fov}(b) and Fig. \ref{scaling}(c), there also exists a trade-off between F/\# and pixel count. If smaller F/\# is preferred to increase light collection ability, smaller image fill factor $\Gamma$ should be used, but will result in reduced pixel count.
\begin{figure}[ht!]
	\centering\includegraphics[width=\textwidth]{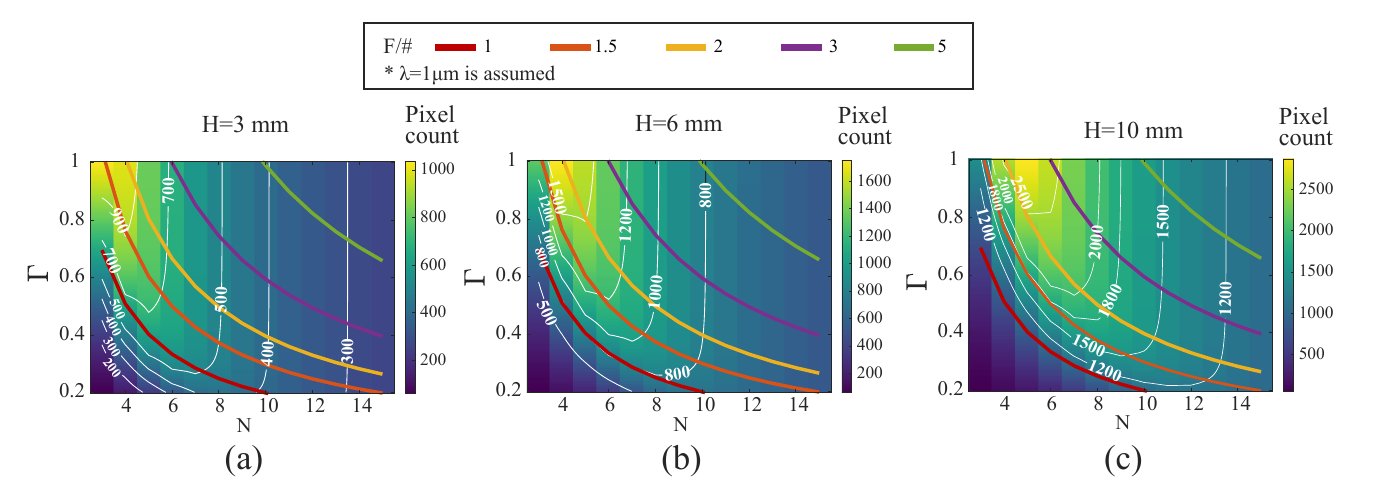}
	\caption{Achievable F/\# and lateral pixel count using different image fill factor $\Gamma$, number of units $N$ and lateral imager height $H$.}
	\label{scaling2}
\end{figure}

Combining Eq. \ref{cnt} and Eq. \ref{F_num2}, the relation of lateral pixel count, $\Gamma$ , $N$ and achievable F/\# are illustrated in Figs. \ref{scaling2}(a-c), where three different imager height conditions (3 mm, 6 mm and 10 mm) are chosen. With these figures, one could choose maximum pixel count or minimum F/\# and get corresponding design parameters. Note that the empirical fitting model shown in Fig. \ref{scaling}(b) is used to evaluate the geometric spot size, thus the exact lateral pixel count may deviate from Fig.  \ref{scaling}(b). Nevertheless, Fig. \ref{scaling}(b) is still a useful reference during design process, which will be shown in the Next section.

\section{Design examples}

\begin{figure}[ht!]
	\centering\includegraphics[width=0.8\linewidth]{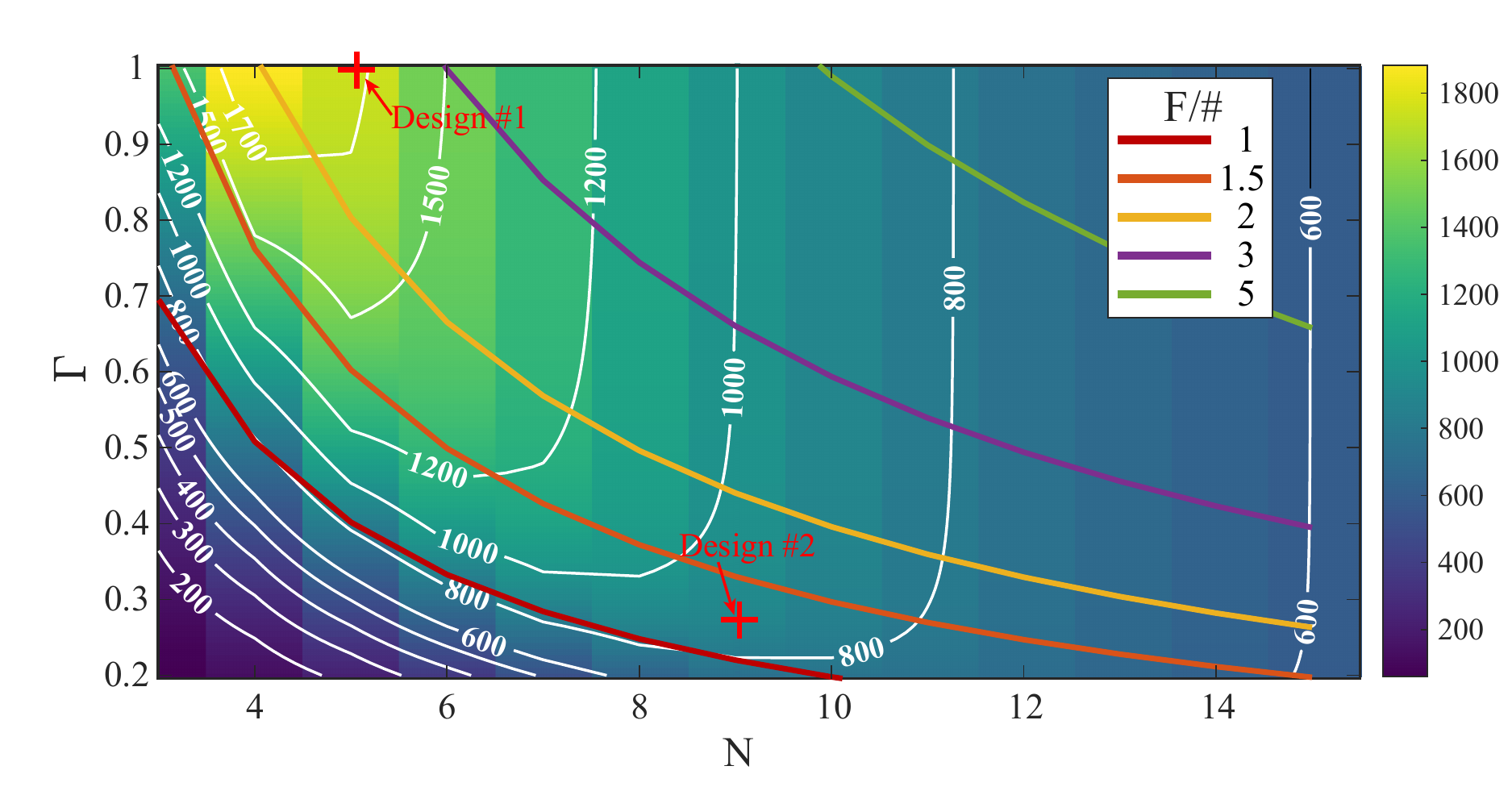}
	\caption{Achievable F/\# and lateral pixel count using different image fill factor $\Gamma$, number of units $N$ and lateral imager height $H$ to fully cover \ang{+-80} FoV. The parameters of two design examples are marked in the figure. $\lambda=0.9\mu$m is assumed.}
	\label{design1}
\end{figure}

Here we use two design examples to demonstrate the superiority of the system and effectiveness of the design method described above. In our design examples, we assume wavelength $\lambda = \SI{0.9}{\um}$ and lateral imager size $H = \SI{10}{\mm}$, which is close to the the size of the modern mobile phone CMOS sensor. The FoV of the system is chosen as $\pm \ang{80}$ considering the feasibility of metagratings. From Eq.\ref{fov_eq}, we then choose the needed number of units $N$ and AOI range $[-\alpha_0,\alpha_0]$ to fully cover the \ang{+-80} FoV.  Feasible parameters for FoV$=\ang{80}$ have been shown in Fig. \ref{fov}. For $N\ge 4$, $\alpha_0 <\ang{15}$ can be fulfilled to ensure the feasibility of the AOI range of metagratings, and we can select larger $N$ to further reduce the AOI range. Considering the factors, we choose $N=5$ and $N=9$ for our two design examples, leading to $\alpha_0=\ang{11.36}$ and $\alpha_0=\ang{6.28}$, respectively. To further determine image fill factor $\Gamma$ and estimate achievable pixel count and F/\# before optical design, we redraw Fig. \ref{scaling2} with newly designated parameters in Fig. \ref{design1}, where the balance between F/\# and resolution can be make. Here we choose $\Gamma = 1$ for design example with $N=5$ and $\Gamma=0.23$ for design example with $N=9$, and we can expect a lateral pixel count around 1700 and 800, respectively. From Fig. \ref{fov}(b) or Eq. \ref{F_num2}, the corresponding F number for two design examples are roughly 2.5 and 1.0, respectively. Note that effective lateral pixel count and F/\# estimated here are only a rough prediction using empirical formula Eq. \ref{delta_g}. The exact values will be given after optical design procedure. The metagrating parameters $g_{m,i} = m_i \lambda/\Lambda_i$ for these two designs are thus decided by Eq.\ref{grating_lines}, which are shown in Table.\ref{grating_tb}.
\begin{table}[ht!]
	\centering
	\begin{threeparttable}[b]
		\caption{Metagrating parameters}
		\begin{tabular}{llccccc}
			\hline
			\multicolumn{2}{r}{Unit \# $^\text{a}$, i} & 0 & \num{+-1}     & \num{+-2}     & \num{+-3}  & \num{+-4} \\ 
			\hline
			\multirow{2}{4em}{Design \#1} &
			Grating Param.$^\text{b}$, $g_{m,i}$ & - & 0.219 & 0.438 &  &  \\ 
			\cline{2-7}
			& Grating Period $^\text{c}$ ($\mu$m) & - & 4.11 & 2.06 & & \\
			\hline
			\multirow{2}{4em}{Design \#2} &
			Grating Param.$^\text{b}$, $g_{m,i}$ & - & 0.219 & 0.438 & 0.657 & 0.875 \\ 
			\cline{2-7}
			& Grating Period $^\text{c}$ ($\mu$m) & - & 4.11 & 2.06 & 1.37 & 1.03 \\
			\hline
		\end{tabular}
		\begin{tablenotes}
			\item [a] $i$ is counted from the center of the array to its top edge and thus ranges from 0 to $(N-1)/2$ for odd-number units. Negative $i$ means counting from the center of the array to its bottom edge.
			\item [b] $g_{m,i} = m_i \lambda/\Lambda_i$ for i-th metagrating; ''-'' means that grating is not needed for the certain unit.
			\item [c] Assuming the operating order of metagratings is $m=1$
		\end{tablenotes}
		\label{grating_tb}
	\end{threeparttable}
\end{table}

\begin{figure}[ht!]
	\centering\includegraphics[width =\textwidth]{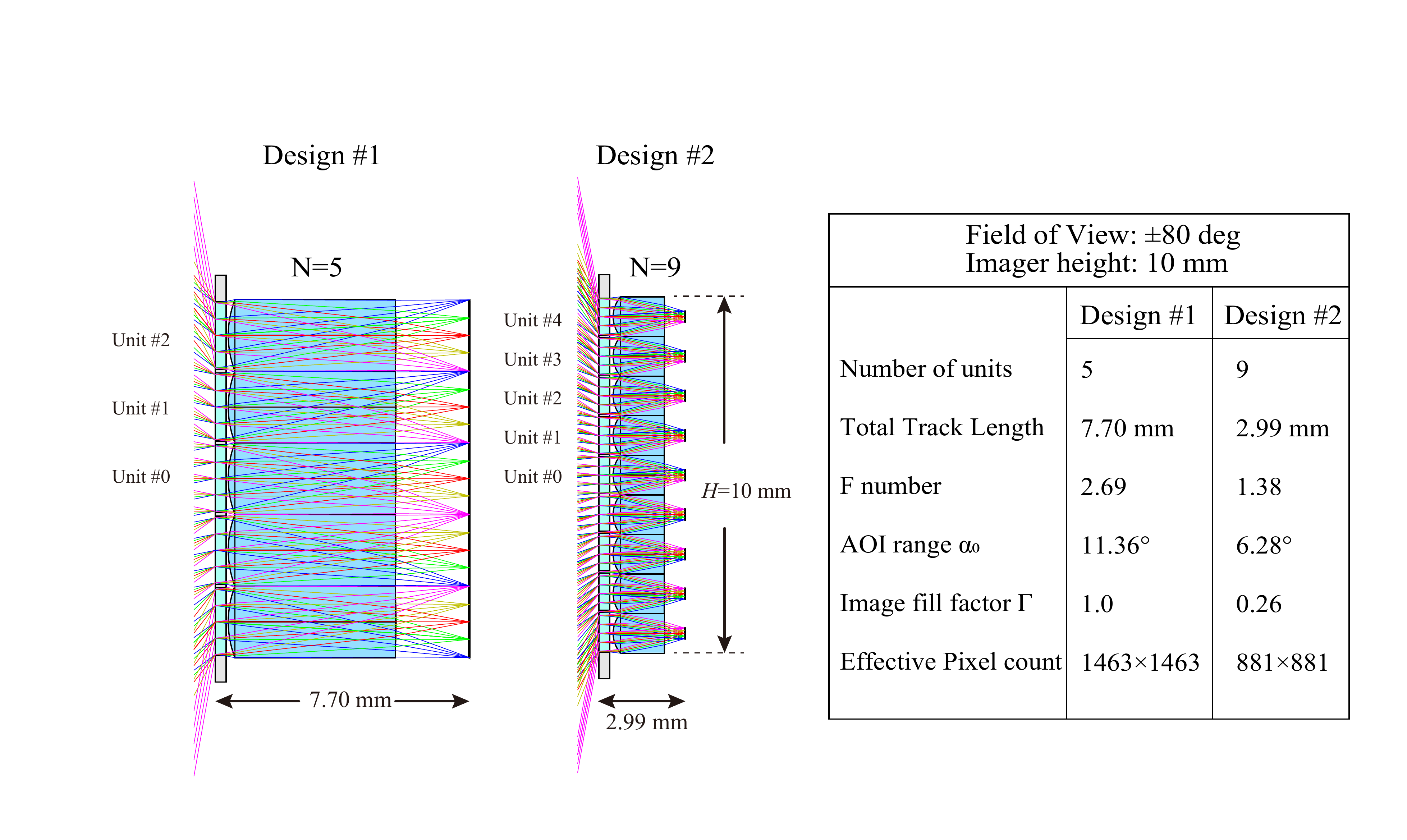}
	\caption{Designed optical systems with ray tracing result and system specification.}
	\label{zmx}
\end{figure}

The designed systems with ray tracing results and system specification are shown in Fig. \ref{zmx}. We can see the similarity among the units, which is brought by the metagratings. The introduce of metagratings is able to convert highly slanted sub-FoV into near-axis sub-FoV, so that each lenslet operates with identical near-axis FoV and becomes identical with each other. As a result, optical design for each lenslet becomes easy because one can only design a single lenslet unit and apply the same parameter to other units. Each lenslet uses 4-order aspherical surfaces to reduce spherical aberration. The material of the lenslet array is chosen as optical plastic E48R, enabling the stantard injection molding fabrication process. The material of the substrate of the metagrating array is fused silica. As we do not discuss the design of metagratings in this paper, we have modeled them as diffraction gratings with a single operating order. The first design with fewer units has a relatively larger total track length, which is \SI{7.70}{\mm}, but is still very compact. Also, the system possesses high imaging resolution up to 1463 lateral pixel count, which is lower than the prediction given by Fig. \ref{design1} due to the fitting error of geometric spot diameter shown in Fig. \ref{scaling}. Nevertheless, the design method is still effective. The second design has a more compact structure with a total track length of only \SI{2.99}{\mm}, but with the cost of lower imaging resolution (881 lateral pixel count) and smaller entrance pupil.

\begin{figure}
	\centering
	\includegraphics[width=0.85\textwidth]{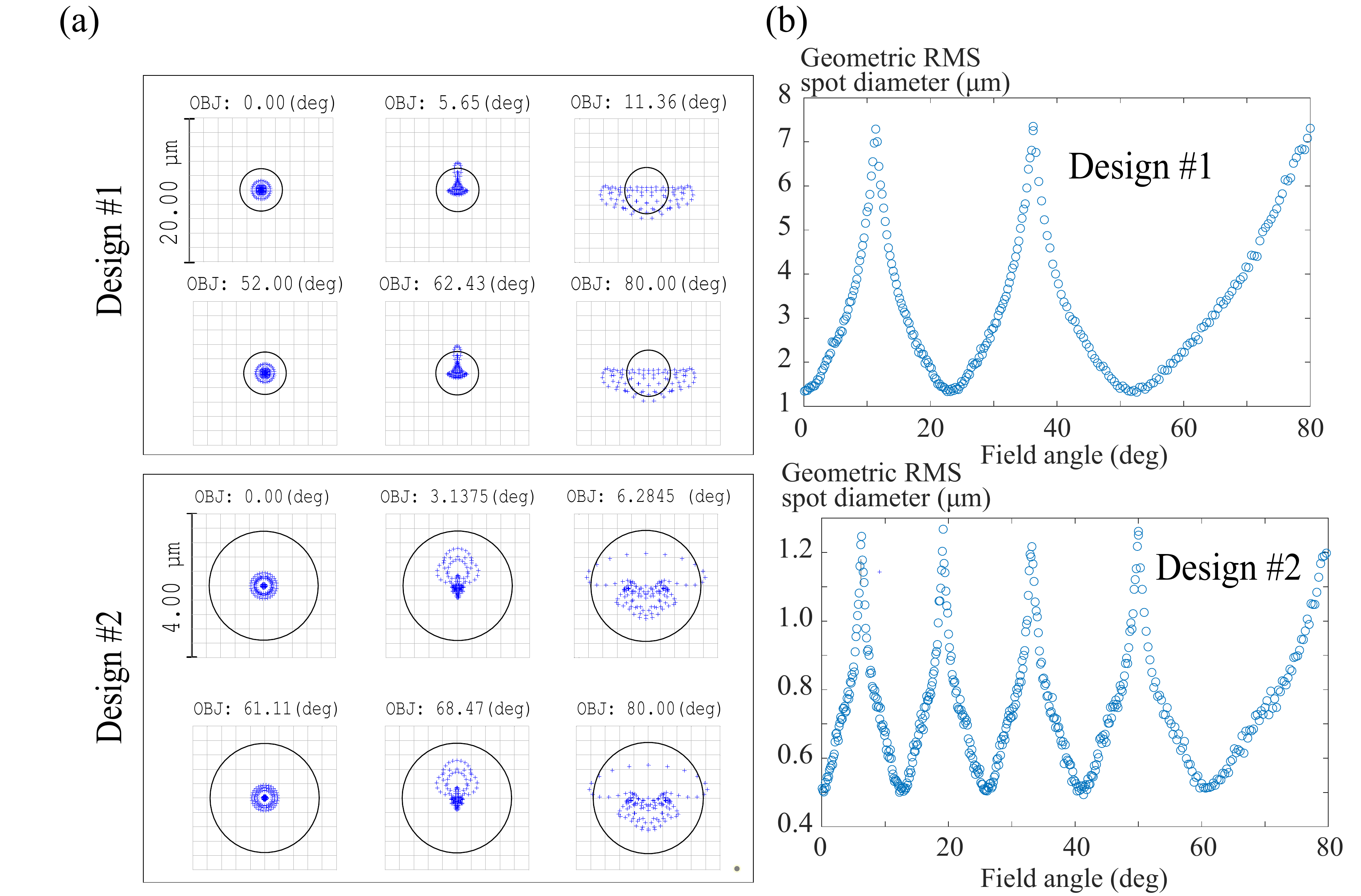}
	\caption{(a) Spot diagram and (b) root-mean-square geometric spot diameters across the whole FoV. Diffraction limit (Airy disk) is shown as black circles in (a).}
	\label{spt}
\end{figure}

To evaluate the imaging quality of the two design examples, spot diagrams are shown in Fig. \ref{spt}(a) and root-mean-square geometric spot diameters across whole FoV are shown in Fig. \ref{spt} (b). The diffraction limited focal spot, i.e., Airy disk is plotted as black circles in Fig. \ref{spt} (a) for reference. From the spot diagram we can see that geometric aberration becomes quasi-periodic rather than monotonic increasing as field angle increase, showing that the proposed design can effectively eliminate off-axis aberration. The spot diagrams also show that both systems are diffraction limited within the whole field, which is different from geometric aberration limited conventional fish-eye lens. This is because geometric aberration scales with the size of lenses but diffraction limit is decided by F/\#, which does not change when the size of lenses scales down or up \cite{Lohmann1989}.

\begin{figure}[ht!]
	\centering
	\includegraphics[width=0.8\textwidth]{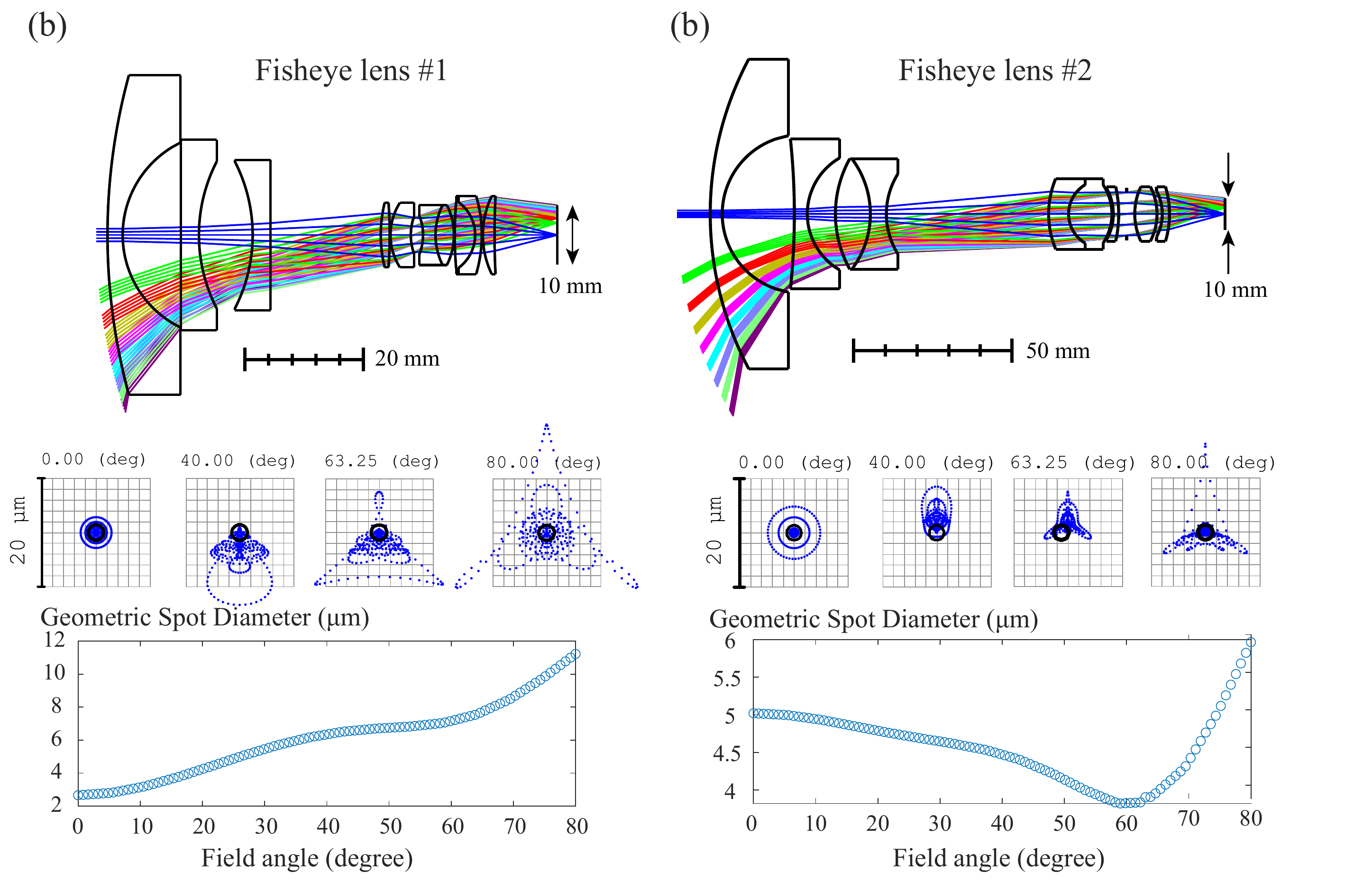}
	\caption{Spot diagram of two typical conventional fish-eye lens (adopted from Zebase library). They are all scaled so that the imager size is also \SI{10}{\mm}}
	\label{fisheye}
\end{figure}

Next, to further assess the effectiveness of the proposed solution, we compare the performance of the design examples with two typical conventional fish-eye lenses which are adopted from Zebase library. The two fish-eye lenses are both scaled to the same imager size with our design, i.e. \SI{10}{\mm}. We show design parameters, structures and spot diagrams of the two lenses in Fig. \ref{fisheye}. Now we can clearly see that both the two lenses have a strong off-axis aberration, and the spot sizes are much larger than our design examples. The fact means our design possesses roughly the comparable or even superior imaging quality with conventional designs. However, our proposed systems are far more compact than conventional fish-eye designs. We see that even for the first longer design example, the total track length is only \num{1/10} and \num{1/20} of the two conventional fish-eye designs, respectively. 

\section{Discussion}

\begin{figure}[ht!]
	\centering
	\includegraphics[width=0.8\textwidth]{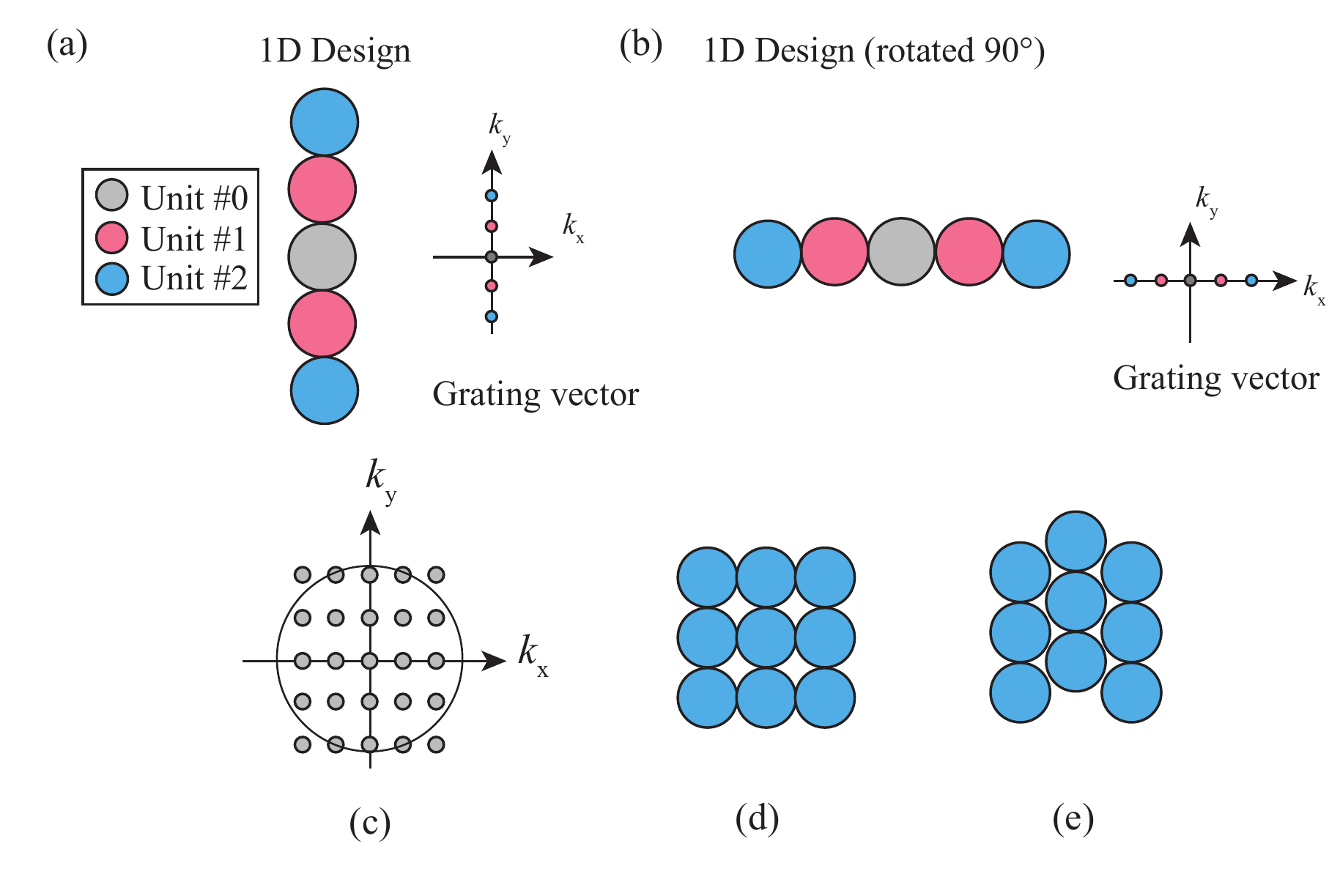}
	\caption{2D stitching method. }
	\label{stitch}
\end{figure}

In above analysis, we only consider the design process in one dimension which is denoted as y-axis. The final two-dimensional arrangment can be obtained by repeating the design process in another dimension and apply an appropriate stitching strategy. As all lenslet units are identical, the lenslet array can be obtained by simply repeating a single unit. However, metagrating parameters varies along both dimensions. Grating vector along each axis is defined as
\begin{equation}
    k_x = \frac{m}{\Lambda_x},\;\; k_y = \frac{m}{\Lambda_y},
\end{equation}
where $k_y$ is the same as one-dimensional case, and grating period $\Lambda_y $ along y-axis is given by Eq. \ref{grating_lines}. From grating equation, grating vector denotes the power and direction of deflection. For two-dimensional case, the desired metagrating units with a certain length of grating vector $|\mathbf{k}|$ can be obtained by rotation. For example, a 90-degree rotation of metagrating unit can generate a grating with a grating vector along the orthogonal direction, as shown in Fig. \ref{stitch} (a) and (b). Thus, the grating parameters of metagrating units can be obtained, which is illustrated in Fig. \ref{stitch}(c) and (d). Finally, a stitching strategy should be chosen to map each unit in k-space onto a certain position. For example, square arrangement and hexagonal arrangement can be applied, which are shown in Fig. \ref{stitch}(d) and (e).

From Fig. \ref{spt}, these two design examples are diffraction limited but not geometric aberration limited as conventional fish-eye lens. For such systems, small F/\# is critical to increase resolution and thus improve imaging quality. As we have mentioned in Section. 2, if we do not want to have an increased AOI range for metagratings, which is challenging, further reducing F/\# means a larger entrance aperture. When entrance apertures becomes larger for each unit, we must allow of a larger lenslet array than the image size, leading to tilted chief rays very similar to the case of conventional multi-aperture lens shown in Fig. \ref{gallery}. Assuming that the entrance pupil diameter for the modified structure is $M$ times the length of the initial design, the F/\# of the modified design will reduce $M$ times. As a result, the design for each unit is no longer similar and requires to be optimized separately. Similar to the normal milti-aperture lens in Fig. \ref{gallery}, the F/\# reduction multiplier $M$ is finally limited by the difficulty of optical design.

In the proposed design, chromatic performance is not considered because such metagratings used in the system are generally highly dispersive especially for large incident field angles. This is also the main drawback of our proposed method compared to fully lens-based solutions. However, in applications such as LiDAR, night-vision surveillance camera and microscopy, campact monochromatic lenses with large FoV is highly demanded and the proposed system will be a potential solution. Besides, various achromatic metasurfaces and metagratings have been reported and such device can operate as a broadband deflector without spatial dispersion \cite{Aieta2015,Khorasaninejad2015}. With such devices, the proposed design is a promising solution to achieve large-FoV and ultra-compact achromatic lens.

\section{Conclusion}

In this paper, we proposed a very compact fish-eye lens design with a planer focal plane which contains only two planar elements, i.e. a metagrating array and a lenslet array. The proposed design method can effectively eliminate field-angle dependent aberration and thus possess superior imaging quality within a large FoV. A systematic design procedure of the lens system is provided. Two design examples with \ang{+-80} FoV are exhibited, which show a similar imaging performance with two typical conventional fish-eye lenses with more than 8 elements. Meanwhile, the total axial length of the design is an order of magnitude smaller than that of the conventional fish-eye cases. The proposed structure and corresponding design method  has the potential to achieve ultrathin or wafer-level fisheye camera. Besides, it expands the field of view of miniaturized multi-aperture lens from less than \ang{+-30} to fish-eye regime that can benefit various applications.

\section*{Funding} National Natural Science Foundation of China (61875104).

\section*{Disclosures} The authors declare no conflicts of interest.


\bibliography{multi-proj}

\end{document}